\documentclass[final,5p,times,twocolumn]{elsarticle}
\usepackage{amsmath,amssymb,mathtools,bm}
\usepackage[T1]{fontenc}
\usepackage[utf8]{inputenc}
\usepackage{xcolor}
\usepackage{hyperref}
\usepackage{ragged2e}
\usepackage{soul}
\usepackage{tikz}
\usetikzlibrary{arrows.meta,decorations.pathmorphing,positioning}
\sethlcolor{yellow}
\soulregister\textit7
\soulregister\textbf7

\newcommand{\DM}{\mathrm{DM}}
\newcommand{\mDM}{\mathrm{mDM}}

\usepackage{ezedits}
\defineEditQuick{SK}{blue}
\defineEditQuick{MG}{red}
\defineEditQuick{Ar}{brown}

\makeatletter
\def\ps@pprintTitle{%
  \def\@oddhead{\footnotesize\itshape\hfill FERMILAB-PUB-26-0828-T}%
  \let\@evenhead\@empty
  \let\@oddfoot\@empty
  \let\@evenfoot\@oddfoot
}
\makeatother

\journal{Physics Letters B}
\biboptions{sort&compress}

\begin{document}
\hypersetup{pageanchor=false}

\begin{frontmatter}

\title{Resonant Cavity Searches for Millicharged Dark Matter}

\author[ifae]{Miguel Gonz\'alez Albert}
\author[normal,nanjing]{Ariel Arza}
\author[ifae,uchicago,fermilab]{Saarik Kalia}
\affiliation[ifae]{address = {Institut de F\'isica d’Altes Energies (IFAE), The Barcelona Institute of Science and Technology, Campus UAB, 08193 Bellaterra (Barcelona), Spain}}
\affiliation[normal]{address = {Department of Physics and Institute of Theoretical Physics, Nanjing Normal University, Nanjing 210023, China}}
\affiliation[nanjing]{address = {Nanjing Key Laboratory of Particle Physics and Astrophysics, Nanjing, 210023, China}}
\affiliation[uchicago]{address = {Enrico Fermi Institute and Leinweber Institute for Theoretical Physics, The University of Chicago, Chicago, IL 60637, USA}}
\affiliation[fermilab]{address = {Theory Division, Fermi National Accelerator Laboratory, Batavia, IL 60510, USA}}

\begin{abstract}
A plethora of resonant cavity experiments have been designed to search for ultralight dark matter (DM) candidates, such as axion or dark-photon DM.  In this work, we show that these same apparatuses can be sensitive to ultralight millicharged dark matter (mDM) with masses $m_\phi\sim1-100\,\mu\mathrm{eV}$.  Similar to axion DM, in the presence of a background magnetic field, mDM acts as an effective current which can excite the modes of a resonant cavity.  Importantly, for a cylindrical cavity, axion DM dominantly couples to the transverse magnetic (TM) modes, whereas mDM dominantly couples to the transverse electric (TE) modes.  We project the sensitivity to mDM that resonant cavity experiments can achieve if they are able to tune and readout their TE modes.  We find that fractional millicharges as small as $\epsilon_m=e_m/e\sim10^{-19}$ can be probed.  If mDM constitutes only a subfraction $f_\mDM \lesssim0.2\%$ of the total DM density, competing astrophysical bounds do not apply.  In this case, resonant cavities can be the most sensitive probes of mDM in this mass range.
\end{abstract}

\end{frontmatter}
\hypersetup{pageanchor=true}

\section{Introduction}
At present, the fundamental nature of dark matter (DM) remains unknown. Its elusive character and weak interactions motivate a variety of viable candidates, several of which have already been investigated in considerable detail.

Much recent attention has focused on detection of ultralight DM with masses $m_\DM\lesssim10\,\mathrm{eV}$~\cite{Arias:2012az,kimball2022search,Antypas:2022asj}.  This class of candidates, which includes dark-photon DM~\cite{Holdom:1986ag,Nelson:2011sf,Graham:2015rva,Okun:1982xi} and axionlike-particle DM (or simply axion DM)~\cite{Preskill:1982cy,Abbott:1982af,Dine:1982ah,Svrcek:2006yi,Arvanitaki:2009fg}, behave as classical waves~\cite{cheong2024} due to their high phase-space occupancy.  If they couple to electromagnetism (EM), they act as an ``effective current" that sources Maxwell's equations, which can be observed in various EM apparatuses~\cite{ehret2010new,Redondo:2010dp,Horns:2012jf,Chaudhuri:2014dla,Caldwell:2016dcw,Lawson:2019brd,gramolin2021search,Fedderke_2021}.  One of the most mature detection schemes to probe these candidates is the resonant cavity~\cite{sikivie1983,grenet2021grahal,chang2022taseh,adair2022castcapp,quiskamp2025organ,bai2025,carosi2025,Sardo_Infirri_2025}.  In the presence of a large background magnetic field, axion DM can convert into photons inside a resonant cavity, which is readout as excess power by a low-noise receiver chain.  Resonant cavity experiments are most sensitive to DM whose mass matches the resonant frequency of the cavity mode, and this resonant frequency can typically be tuned mechanically.  These experiments achieve sensitivity to tiny DM couplings by leveraging a strong magnetic field and a high cavity quality factor $Q$.

Millicharged particles (mCPs), i.e., particles carrying an electric charge $e_m=\epsilon_m e$ much smaller than the electron charge $e$, constitute another well-motivated DM candidate \cite{dimopoulos1990,derujula1990,feldman2007,mcdermott2011,cline2012}. Here $e_m$ denotes the physical electromagnetic coupling in the covariant derivative, while $\epsilon_m\equiv e_m/e$ is the dimensionless charge fraction conventionally quoted. They arise naturally in scenarios with kinetic mixing between the visible and dark sectors \cite{feldman2007,Holdom:1986ag,goldberg1986,izaguirre2015,cheung2007,feng2024} or in models where these particles are directly charged under the $U(1)_Y$ of the Standard Model \cite{wen1985}. They are also predicted in string-theory compactifications \cite{wen1985,shiu2013,feng2014,burgess2008,goodsell2009,cicoli2011} and grand-unified theories \cite{pati1973,georgi1975}.  Light mCPs can be produced via the misalignment mechanism with the correct abundance to make up the entirety of DM~\cite{Bogorad_2021}.

\begin{figure}[t]
    \centering
    \begin{minipage}{0.48\columnwidth}
        \centering
        \includegraphics[width=\linewidth]{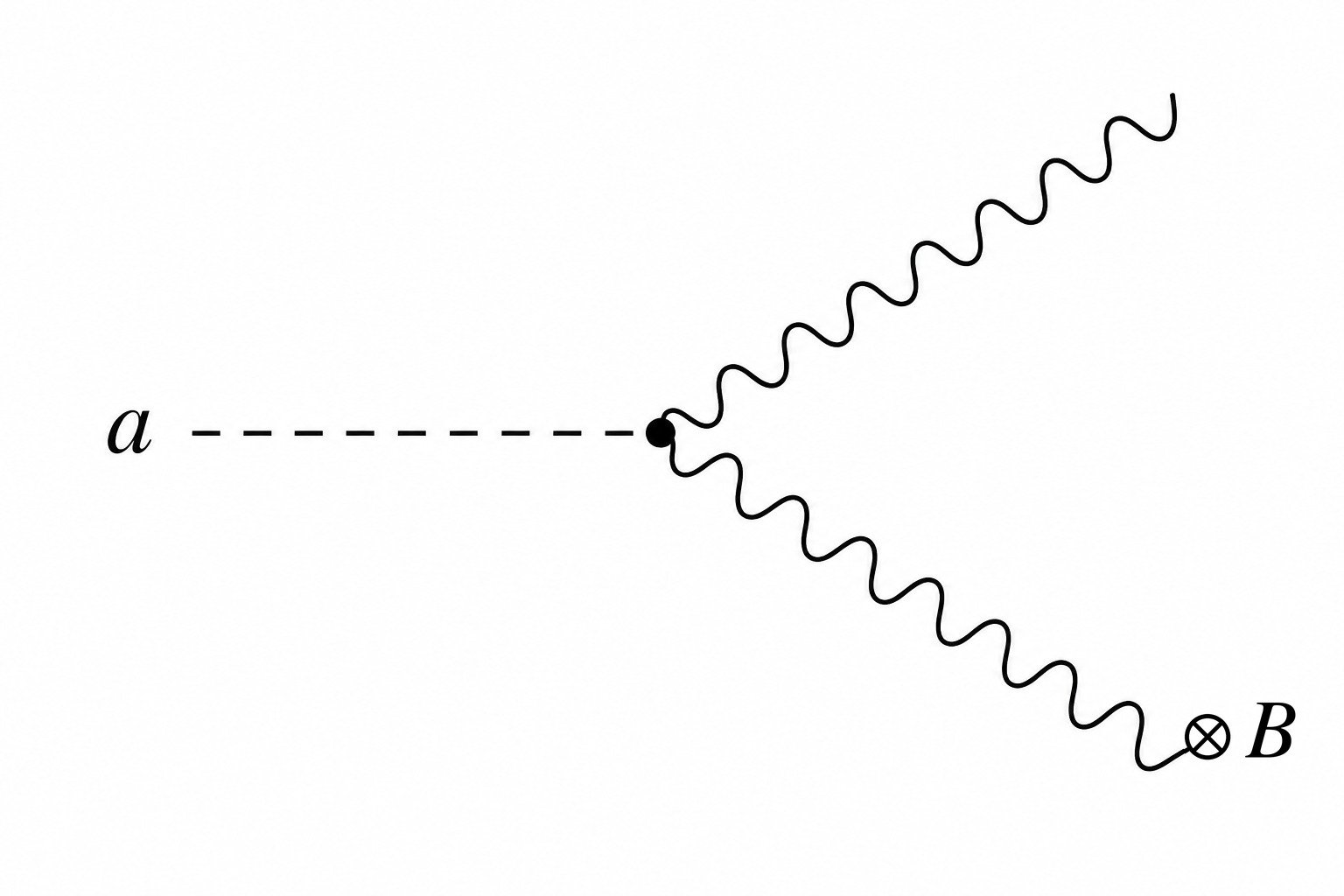}
    \end{minipage}
    \hfill
    \begin{minipage}{0.48\columnwidth}
        \centering
        \includegraphics[width=\linewidth]{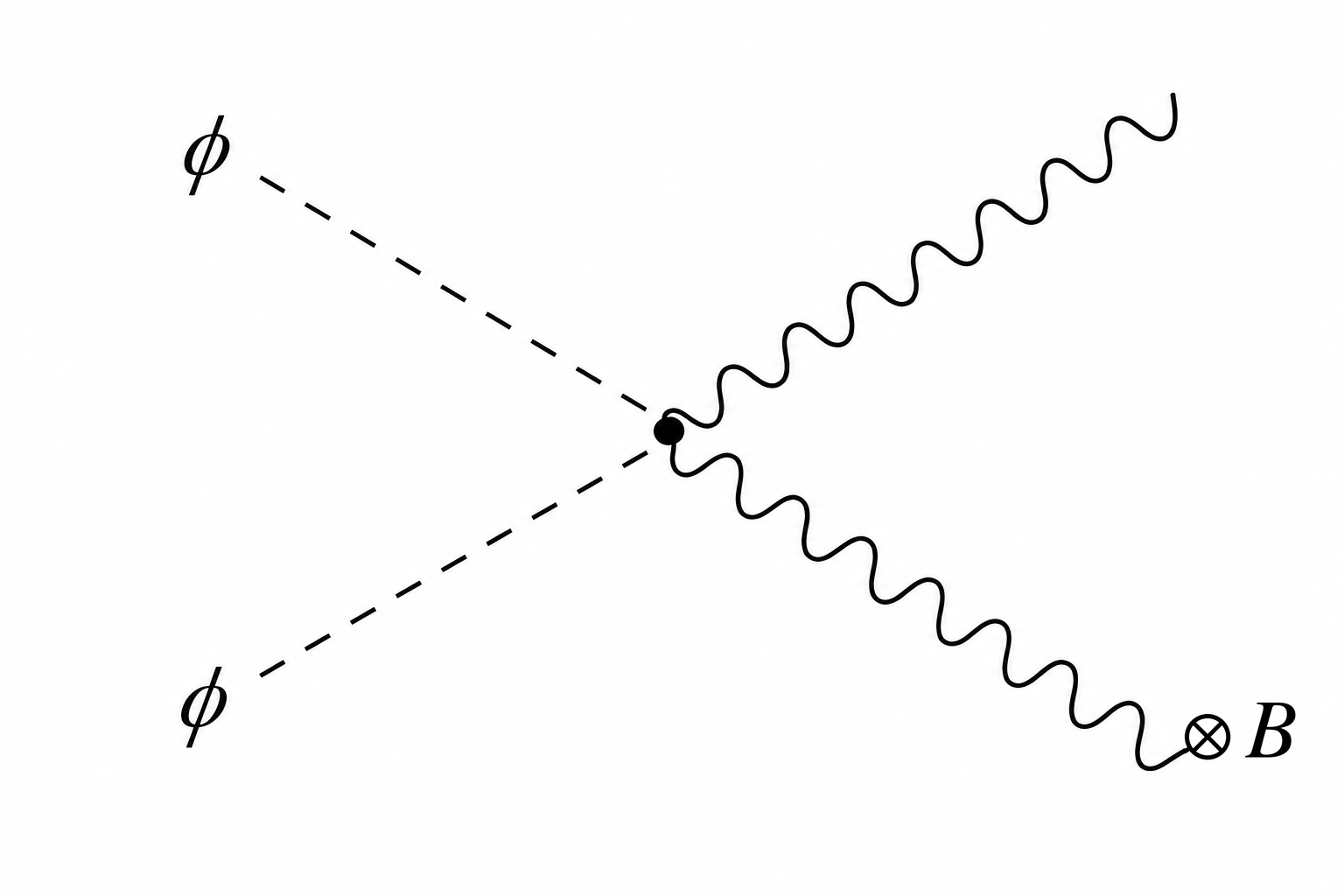}
    \end{minipage}

    \caption{Comparison of the axion--photon and relevant millicharge--photon couplings.}
    \label{fig:couplings}
\end{figure}

In the ultralight regime, millicharged dark matter (mDM) also behaves as a classical wave and can be detected in analogy to EM-coupled axion DM~\cite{arza2025}.  Specifically, in the presence of a background magnetic field, mDM can also act as an effective current present in Maxwell's equations.  Fig.~\ref{fig:couplings} shows how the two-photon coupling of scalar mDM allows for conversion of two mCPs into a photon inside a background magnetic field, analogous to the usual axion--photon conversion. This connection implies that experiments designed to detect axion DM should also exhibit sensitivity to mDM.  In this work, we demonstrate this analogy for resonant cavity experiments and project their sensitivity to mDM.  Importantly, we find that mDM dominantly couples to transverse electric (TE) cavity modes, in contrast to the transverse magnetic (TM) modes that axion DM couples to. As a result many cavity experiments are not sensitive to mDM in their current configuration.  Nevertheless, if existing experiments are able to readout and tune their TE modes, they can become the most sensitive probes of subcomponent mDM in the $\mu\mathrm{eV}$ mass range.

This work is structured as follows.  In Sec.~\ref{sec:cavities}, we review the basic properties of resonant cavities.  In Sec.~\ref{sec:mcp}, we derive the effective current description for mDM.  In Sec.~\ref{sec:readout}, we compute the overlap of this current with the relevant cavity modes and derive the power that a resonant cavity would read out.  In Sec.~\ref{sec:sensitivity}, we utilize existing axion limits from cavity experiments to project sensitivities to mDM.  In Sec.~\ref{sec:constraints}, we discuss competing constraints from astrophysics and cosmology.  In Sec.~\ref{sec:conclusion}, we conclude.  In the Appendix, we discuss how the mDM can be distorted by background magnetic fields and derive the necessary conditions to avoid this distortion.

\section{Resonant Cavities}
\label{sec:cavities}

In this class of experiments, a high-quality-factor microwave cavity is placed inside a strong static magnetic field and operated at cryogenic temperature. The cavity acts as a narrowband resonator: only electromagnetic modes with frequencies close to the signal frequency are efficiently enhanced. The resonant frequency is tuned mechanically, for example by moving conducting or dielectric tuning rods, so that the experiment scans a range of possible dark-matter masses.  In the standard axion case, the static magnetic field converts the axion field into a weak microwave signal through the axion--photon coupling \cite{graham2015,sikivie1983}. The expected signal power is proportional to the magnetic-field strength, the cavity volume, the loaded quality factor, and a form factor that measures the overlap between the signal current and the electromagnetic mode. In practice, the cavity is weakly coupled to a readout antenna. Any resonantly enhanced microwave power extracted through this port is amplified by a low-noise receiver chain and searched for as a narrow spectral excess over the thermal and amplifier noise.

ADMX provides the canonical large-volume implementation of this idea, optimized for relatively low microwave frequencies and using a strong solenoidal field, a tunable cylindrical cavity, and ultra-low-noise microwave amplification \cite{asztalos2010,khatiwada2021,du2018,braine2020,bartram2021,goodman2025,carosi2025}. Other haloscope experiments follow the same basic principle but explore different frequency ranges and detector technologies. HAYSTAC uses a smaller cavity at higher frequencies and has implemented quantum-noise-reduction techniques, such as squeezed-state receivers, to improve sensitivity \cite{brubaker2017,zhong2018,backes2021,jewell2023,bai2025}. ADMX Sidecar targets higher-frequency windows with a smaller auxiliary cavity \cite{boutan2018,bartram2023}. QUAX has reported searches with conventional, superconducting, and high-$Q$ dielectric cavities, for which the mode-specific form factor must be calculated rather than inferred from the empty-cylinder result used below \cite{alesini2019quax,alesini2021quax,alesini2022quax,divora2023quax,rettaroli2024quaxlnf}. Other examples include the CAPP cavity program, which explores high-field implementations, the TASEH search near $19.6\,\mu\mathrm{eV}$, and the GrAHal platform and prototype \cite{lee2020capp8tb,yoon2022capp18t,chang2022taseh,grenet2021grahal}.

Experiments hosted in the CAST dipole illustrate another way to extend the reach of cavity searches. CAST-CAPP used a phase-matched array of four rectangular cavities, while CAST-RADES demonstrated a filter-like resonator made of five inductively coupled sub-cavities; both approaches address the loss of useful volume that otherwise limits conventional cavities at higher frequencies \cite{adair2022castcapp,alvarezmelcon2021castrades}. These examples show that the haloscope community is not tied to a single cavity geometry: magnet configuration, resonator architecture, mode structure, and readout are all being developed together.

Several proposed designs further modify the resonator itself. A photonic-crystal haloscope uses a periodic array of dielectric elements inside a conducting cavity to maintain a sizable active volume and high quality factor at higher frequencies \cite{bae2023photonic}. A different, hybrid-mode approach to heterodyne axion detection uses two cavity modes: the axion field drives transitions between a pumped mode and a signal mode, with the frequency difference setting the target axion frequency. A proof-of-principle cavity supporting nearly degenerate, linearly polarized hybrid modes has been fabricated and characterized, including tuning of the mode separation and suppression of unwanted cross-coupling; this prototype is not itself a dark-matter search \cite{li2026hybrid}. The hybrid-mode concept targets frequencies well below the $\mu$eV regime considered here and is distinct from the single-mode axion limits used in our analysis. Applying either these designs or other nonstandard resonators to mDM would require calculating the corresponding scalar-induced source, mode overlaps, and readout response for that specific geometry.

Split-cavity and multicell designs such as CAST-CAPP and CAST-RADES may also require a dedicated reassessment for mDM. In the uniform axial magnetic-field configuration considered here, Eq.~\eqref{eq:Jnorm} shows that the effective source current is azimuthal, circulating around the cavity axis. Conducting partitions can change the vector-potential profile and cavity boundary conditions, and may disrupt the spatial overlap or phase coherence of this source across cells. They do not, however, block the mDM field itself, and each cell may still contain a local source. The net signal therefore depends on the detailed field and mode profiles of the segmented geometry and must be calculated rather than inferred directly from a conventional single-cavity limit.

The common feature of these experiments is therefore not a specific cavity size or frequency range, but the combination of a strong static magnetic field, a tunable high-$Q$ resonator, and a low-noise microwave readout. This makes haloscopes naturally attractive for mDM as well. In the mDM case, the signal is not the usual axion--photon conversion, but rather an effective electromagnetic current induced by the charged scalar field in the external vector potential. Once this current is projected onto the appropriate cavity mode, the same resonant enhancement and readout strategy can be used. Existing axion-haloscope data can therefore be recast as sensitivity to mDM, provided that the correct mode overlap, frequency relation, and density assumptions are used.

\section{mCP Effective Current}
\label{sec:mcp}
The starting point of the calculation is the effective Lagrangian:

\begin{equation}
\mathcal{L}_{\mathrm{eff}} = (D_\mu\phi)(D^\mu\phi)^* - m^2_\phi|\phi|^2 - \frac{1}{4}F_{\mu\nu}F^{\mu\nu}
\label{eq:Leff}
\end{equation}
where $\phi$ is the mDM field, $F_{\mu\nu} = \partial_\mu A_\nu - \partial_\nu A_\mu$ is the electromagnetic field-strength tensor, and the physical millicharge $e_m=\epsilon_m e$ is introduced through the covariant derivative $D_{\mu} = \partial_\mu + ie_mA_\mu$.

Since the occupation number of mCPs with masses below the eV scale is large, the mCP field can be treated as a classical wave \cite{kimball2022search,cheong2024}, and the induced electromagnetic signal can be derived from the modified Maxwell equations. The equations of motion derived from the Lagrangian in \eqref{eq:Leff} are:

\begin{equation}
 (\Box +m_\phi^2)\phi = e_m^2 A_\mu A^\mu \phi - ie_m\partial_\mu A^\mu \phi - 2ie_m A^\mu (\partial_\mu \phi)
 \label{eq:eom}
\end{equation}
\begin{equation}
\partial_\nu F^{\mu\nu} = \mathcal{J}_{\mathrm{eff}}^\mu = i e_m\big[(\partial^\mu \phi)\phi^* - \phi(\partial^\mu\phi)^*\big] - 2 e_m^2 A^\mu |\phi|^2.
\label{eq:J}
\end{equation}
In a manner analogous to axion haloscope searches \cite{sikivie1983}, the oscillating mDM density can source an electromagnetic current in the presence of a static magnetic field. This induced current can therefore be exploited by resonant-cavity experiments.

For a charged scalar field it is useful to write
\begin{equation}
\phi(t,\mathbf{x})=\Phi(t,\mathbf{x})e^{i\alpha(t,\mathbf{x})}.
\end{equation}
where $\Phi$ and $\alpha$ are real. Then Eq.~\eqref{eq:J} can be rewritten as
\begin{equation}
\mathcal{J}_{\mathrm{eff}}^\mu
= -2 e_m |\Phi|^2\left(\partial^\mu\alpha+e_m A^\mu\right),
\label{eq:Jgauge}
\end{equation}
which depends on the gauge-invariant combination $\partial^\mu\alpha+e_m A^\mu$. Far from electromagnetic fields, the dark-matter configuration is assumed to be electrically neutral on average. This implies that, far from the experiment, its effective charge density must vanish,
\begin{equation}
\mathcal{J}_{\mathrm{eff}}^0
= -2 e_m |\Phi|^2\left(\partial^0\alpha+e_m A^0\right)=0.
\end{equation}
Taking temporal Coulomb gauge, $A_0=0$ and $\nabla\cdot\mathbf A=0$, this condition implies $\partial_0\alpha=0$.  The spatial phase gradients are velocity-suppressed for nonrelativistic DM and are neglected at leading order, $\partial_i\alpha\simeq 0$.   Thus $\alpha$ is a constant to this order, and the remaining constant phase can be set to zero.

Far from the experiment, the mDM is then just described by an oscillating amplitude
\begin{equation}
\Phi(t)=\frac{\sqrt{\rho_\mDM}}{m_\phi}\cos(m_\phi t),
\label{eq:phi_cosine}
\end{equation}
where $\rho_\mDM$ is the local density of the millicharged component.  At the location of the experiment, the mDM can be distorted from Eq.~\eqref{eq:phi_cosine} by magnetic fields, e.g. from the experiment itself or from the Earth. The size of the distortion is controlled by the ratio between the magnetic-field length scale and the mDM de Broglie wavelength, and therefore the two cases fall into different regimes. For the laboratory magnetic field, the characteristic length scale is much smaller than the de Broglie wavelength, while for the geomagnetic field it is much larger. We find that the distortion remains negligible for the most sensitive parameter regions considered in this work.  In particular, the condition obtained for the cavity magnetic field is
\begin{equation}
e_m \lesssim 7\times10^{-16} ,
\end{equation}
and for the geomagnetic field is
\begin{equation}
e_m \lesssim 10^{-15}
\left(\frac{m_\phi}{3\,\mu{\rm eV}}\right)^2 ,
\end{equation}
both of which are above the smallest charges probed by the resonant-cavity searches. We therefore neglect propagation-induced distortions in the main text and use Eq.~\eqref{eq:phi_cosine} throughout. The corresponding calculation, including the separate treatment of the $kr\ll1$ and $kr\gg1$ regimes, is given in the Appendix. 

With this assumption, the resulting time dependence is
\begin{equation}
\left|\Phi(t)\right|^2
=
\frac{\rho_\mDM}{2m_\phi^2}\bigl[1+\cos(\omega t)\bigr],
\qquad
\omega = 2m_\phi.
\label{eq:phi}
\end{equation}
The first term is static and does not contribute to the narrowband resonant signal considered here. In what follows, we retain only the oscillating component. In symmetric gauge, the uniform laboratory magnetic field takes the form
\begin{equation}
\mathbf A=\frac{B_0 r}2 \hat{\theta},
\end{equation}
where $B_0$ is the external magnetic field and $r$ is the radial coordinate. The oscillating current can then be written as
\begin{equation}
\mathbf{J}_{\mathrm{eff}}(t,\mathbf{x})
=
\mathbf{J}_0(\mathbf{x})\cos(\omega t),
\label{eq:Jeff}
\end{equation}
with amplitude

\begin{equation}
\mathbf{J}_0(\mathbf{x})
=
-\frac{e_m^2B_0a\rho_\DM}{2m_\phi^2}\hat{j},
\qquad
\hat{j}=\frac{r}{a}\hat{\theta}.
\label{eq:Jnorm}
\end{equation}

\section{Readout Power}
\label{sec:readout}
To calculate the readout power induced by the mDM field, denoted by $P_\mDM$, we first compute the total stored energy $U$ and the generated power $P_{\mathrm{gen}}$:

\begin{equation}
P_{\mathrm{gen}}= \frac{\omega_n}{Q_L}U,
\qquad
U = W_{en}+W_{mn}.
\label{eq:PU}
\end{equation}
where $Q_L$ is the loaded quality factor of the cavity, and $W_{en}$ and $W_{mn}$ are the stored electric and magnetic energies, respectively.

The readout quality factor $Q_r$ is related to the loaded and unloaded quality factors by
\begin{equation}
\frac{1}{Q_L}=\frac{1}{Q_0}+\frac{1}{Q_r},
\qquad
\beta=\frac{Q_0}{Q_r},
\qquad
Q_L=\frac{Q_0}{1+\beta}.
\label{eq:Qrelations}
\end{equation}
Using these relations, the power extracted through the readout port is therefore
\begin{equation}
P_{\mathrm{out}}=\frac{\omega_n}{Q_r}U
=\frac{\beta}{1+\beta}P_{\mathrm{gen}}.
\label{eq:Pout}
\end{equation}
Here $\beta$ is the antenna coupling coefficient. In this work, we take $\beta=1$, corresponding to critical coupling, $Q_0=Q_r$.

For the modal analysis of the cosine drive in Eq.~\eqref{eq:Jeff}, it is convenient to use the equivalent complex representation. The total electric field is then written as $\mathbf{E}(t,\mathbf{x}) = \Re\!\left[\sum_n c_n\mathbf{E}_n(\mathbf{x})e^{-i\omega t}\right]$ \cite{hill2009}. Near resonance, we retain only the dominant mode $n$ and neglect off-resonant modes and cross terms. The coefficient $c_n$, which encodes the overlap between the cavity resonant mode and the source amplitude $\mathbf{J}_0$, is defined as

\begin{equation}
c_n
= \frac{-i\omega}{\omega_n^2-\omega^2+i\gamma\omega}
\frac{\int_V \mathbf{E}_n^*\!\cdot \mathbf{J}_0\,dV}{\int_V \lvert\mathbf{E}_n\rvert^2\,dV}.
\end{equation}
At resonance, this expression simplifies by using $\gamma=\omega/Q_L$. Combining this with Eq.~\eqref{eq:Jnorm}, one finds
\begin{equation}
c_n
= \frac{Q_L}{\omega}
\frac{e_m^2B_0a\rho_\DM}{2m_\phi^2}
\frac{\int_V \mathbf{E}_n^*\!\cdot \hat{j}\,dV}{\int_V \lvert\mathbf{E}_n\rvert^2\,dV}.
\end{equation}
At resonance, the stored electric and magnetic energies are equal, so the total stored energy associated with the dominant mode is
\begin{equation}
\begin{aligned}
U_n
&= 2W_{en}
= \mathbf{\varepsilon} \int_V \lvert c_n\mathbf{E}_n\rvert^2\,dV \\
&= \left(\frac{e_m^2B_0Q_La\rho_\DM}{4m_\phi^3}\right)^2
\frac{\lvert\int_V \mathbf{E}_n^*\!\cdot \hat{j}\,dV\rvert^2}{\int_V \lvert\mathbf{E}_n\rvert^2\,dV}.
\end{aligned}
\label{eq:W}
\end{equation}
It is then convenient to define
\begin{equation}
C_\mDM^{(n)} = \frac{1}{V} \frac{\lvert \int_V \mathbf{E}_n^*\!\cdot \hat{j}\,dV\rvert ^2}{\int_V \lvert\mathbf{E}_n\rvert^2\,dV}.
\end{equation}
This factor contains the overlap between the mDM field and the cavity mode. Typically, axion haloscope experiments use the lowest TM mode $\mathrm{TM}_{010}$ \cite{sikivie1983}. However, mDM does not couple to those modes, because their longitudinal electric-field component causes the relevant volume integral to vanish. It is therefore necessary to analyze the TE modes. Using the standard cylindrical-cavity mode expressions of Ref.~\cite{hill2009}, the corresponding component gives a nonzero volume integral only when $n=0$ and $q$ is odd. The lowest allowed resonant mode is therefore $\mathrm{TE}_{011}$.

Finally, using Eqs.~\eqref{eq:PU}, \eqref{eq:Pout}, and \eqref{eq:W}, the readout power induced by millicharged particles can be written as\footnote{For the case where $Q_L$ is bigger than $Q_\mathrm{mDM}\sim10^{-6}$, the replacement $Q_L\rightarrow Q_\mathrm{mDM}$ must be done in Eq. \ref{eq:PmDM}. To get this result, an equivalent treatment as in ref. \cite{Hong:2014vua} was performed.}

\begin{equation}
P_\mDM
=
\left(\frac{\beta}{1+\beta}\right)
e_m^4 B_0^2 Q_L a^2 \frac{\rho_\DM^2}{8m_\phi^5} V C_\mDM,
\label{eq:PmDM}
\end{equation}
where
\begin{equation}
C_\mDM^{(011)} = \frac{1}{V} \frac{\lvert \int_V \mathbf{E}_{011}^{TE*}\!\cdot \hat{j}\,dV\rvert ^2}{\int_V \lvert\mathbf{E}_{011}^{TE}\rvert^2\,dV} = 0.22.
\end{equation}
The same procedure can be used to obtain the readout power generated in a rectangular cavity, using
\begin{equation}
\vec{A}=\frac{B_0}{2}\left[ -(y - \frac{L_x}{2}) \hat{x} + (x - \frac{L_y}{2})\hat{y} \right]
\end{equation}
with $L_x$ and $L_y$ the side lengths of the cavity. One obtains the analogous power formula with a different $C$ factor
\begin{equation}
P_\mDM^R
=
e_m^4 B_0^2 Q_L \frac{\rho_{DM}^2}{8m_\phi^5} V C_\mDM^R.
\label{eq:PmDMR}
\end{equation}
The nonvanishing overlap is restricted to two families of
transverse-electric modes. The first family is TE$_{0np}$, for which the
nonzero transverse index $n$ must be a positive even integer and the
longitudinal index $p$ must be odd. The second family is TE$_{m0p}$, with
$m$ a positive even integer and again $p$ odd. Modes outside these two
families have vanishing overlap because their field profiles are odd under at
least one of the transverse or longitudinal reflections of the cavity. The
lowest modes satisfying these selection rules are therefore TE$_{021}$ and
TE$_{201}$. For TE$_{021}$, the generated power is
\begin{equation}
P_{\mathrm{gen}}^{(021)}
=
\left(\frac{\beta}{1+\beta}\right)\frac{e_m^4 B_0^2 Q_L \rho_{\DM}^2}{2\pi^4m_\phi^5}
V L_y^2,
\label{eq:Pgen021}
\end{equation}
while the TE$_{201}$ result follows from Eq.~\eqref{eq:Pgen021} by replacing
$L_y$ with $L_x$. The power extracted through the readout port is obtained by
multiplying the generated power by $\beta/(1+\beta)$, as in
Eq.~\eqref{eq:Pout}.

\section{Sensitivity}
\label{sec:sensitivity}
Similarly, the readout power induced by axions can be written as
\begin{equation}
P_{\mathrm{axion}}
=
\left(\frac{\beta}{1+\beta}\right) g_{a\gamma\gamma}^2 B_0^2 Q_L \frac{\rho_\DM}{m_a} V C_{\mathrm{axion}},
\label{eq:Paxion}
\end{equation}
where the resonance with the idealized cavity now corresponds to TM modes:
\begin{equation}
C_{\mathrm{axion}}^{(010)} = \frac{1}{V} \frac{\lvert \int_V \mathbf{E}_{010}^{TM*}\!\cdot\hat{z}\,dV\rvert ^2}{\int_V \lvert\mathbf{E}_{010}^{TM}\rvert^2\,dV} = 0.69.
\end{equation}

\begin{figure}[!t]
\centering
\includegraphics[width=0.92\linewidth]{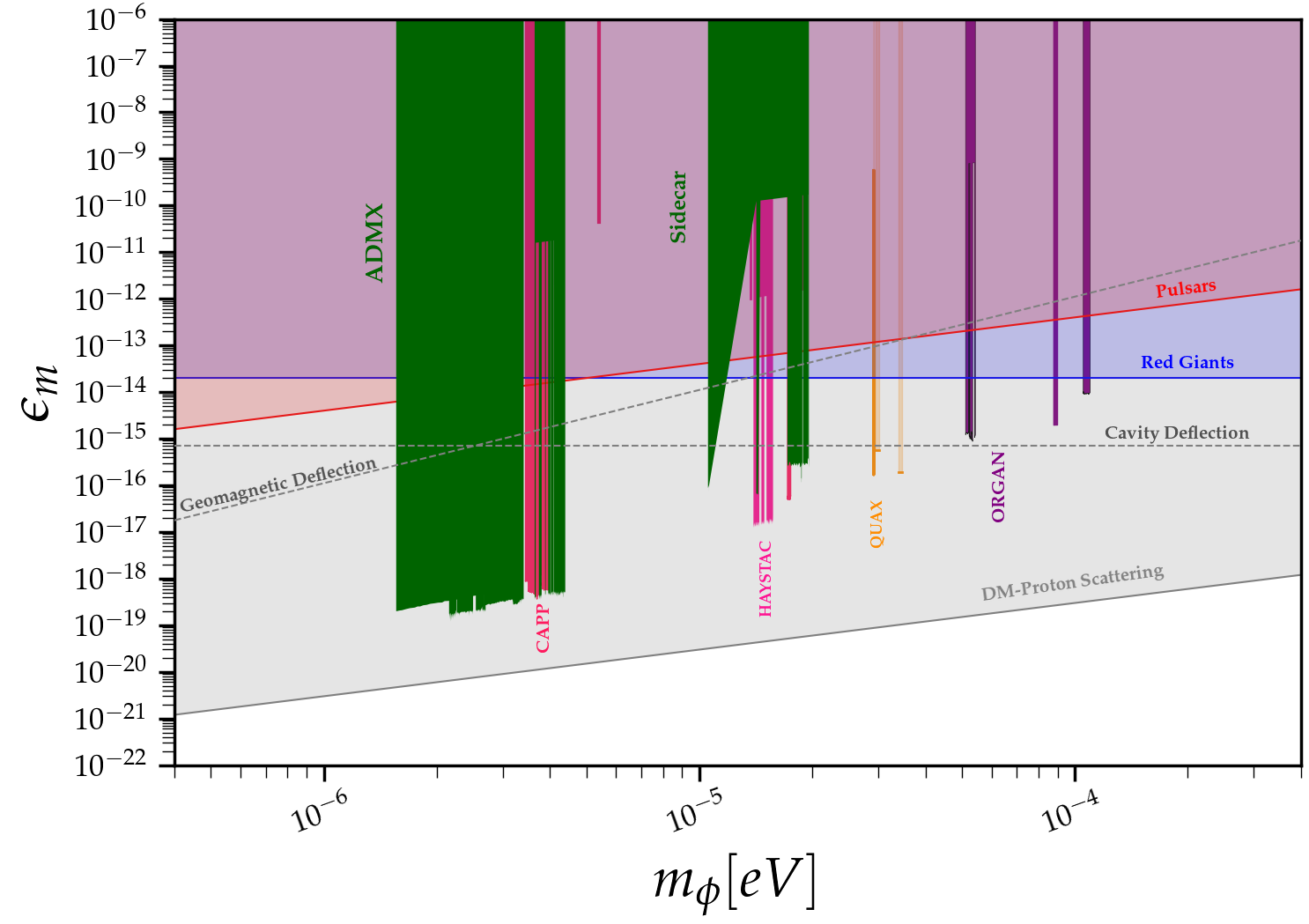}
\caption{Constraints on mDM, quoted in terms of the fractional charge $\epsilon_m=e_m/e$. The red and blue astrophysical regions correspond to pulsar timing and red-giant bounds \cite{caputo2019,davidson2000}, respectively. The labeled vertical haloscope features show the recast derived from Eq.~\eqref{eq:coup} in narrow resonant mass windows for ADMX, Sidecar, CAPP, HAYSTAC, QUAX, and ORGAN, using the published axion limits \cite{asztalos2010,du2018,boutan2018,braine2020,bartram2021,bartram2023,goodman2025,carosi2025,lee2020capp1,yi2023capp6,bae2024capp9,ahn2024cappmax,brubaker2017,zhong2018,backes2021,jewell2023,bai2025,alesini2019quax,alesini2021quax,alesini2022quax,divora2023quax,rettaroli2024quaxlnf,quiskamp2022organ,quiskamp2024organ}. The horizontal mass coordinate is converted according to Eq.~\eqref{eq:freqmap}. Only the QUAX-$a\gamma$ resonant-cavity limits on the axion--photon coupling $g_{a\gamma\gamma}$ are used. The dashed gray lines mark the boundaries above which deflection of the mDM wave by the cavity's or Earth's magnetic field becomes significant, and Eq.~\eqref{eq:phi_cosine} is no longer valid. The gray region indicates the cosmological bound on mDM from scattering with protons in the early universe~\cite{Bogorad_2021,Boddy_2018}.  We also note other astrophysical bounds from large-scale magnetic fields~\cite{stebbins2019,kadota2016} (see text for discussion).  The scattering and magnetic-field bounds do not apply if the mDM constitutes a subfraction $f_\mDM \lesssim0.2\%$ of the total DM density.}
\label{fig:mdm-power}
\end{figure}

For any cavity experiment, one can relate the typical readout power induced by axions in Eq.~\eqref{eq:Paxion} to the readout power induced by mCPs in Eq.~\eqref{eq:PmDM} by establishing a relation between the corresponding couplings. Equation~\eqref{eq:PmDM} is written in terms of the physical charge $e_m$ appearing in the covariant derivative. For comparison with bounds quoted as fractional charges, we use $e_m^4=(\epsilon_m e)^4$. The mapping between axion sensitivity $g_{a\gamma\gamma}$ and millicharge sensitivity $\epsilon_m$ is given by
\begin{equation}
\epsilon_m = \frac{1}{e}\left( g_{a\gamma\gamma}^2 \frac{8m_\phi^5}{a^2\rho_\DM m_a} \frac{C_{\mathrm{axion}}^n}{C_\mDM^{(011)}}\right)^{1/4},
\qquad e_m=e\epsilon_m
\label{eq:coup}
\end{equation}
where $m_{\phi}$ is the mDM mass and $m_a$ is the axion mass used in the comparison. Moreover, $g_{a\gamma\gamma}$ is the axion--photon coupling, which in QCD axion models is given by $g_{a\gamma\gamma}=\frac{\alpha g_\gamma}{\pi f_a}$, where $f_a$ is the axion decay constant and $g_\gamma$ is a model-dependent coefficient ($-0.97$ in the KSVZ model and $0.36$ in the DFSZ model) \cite{graham2015}. The factor $C_\mDM^{(011)}$ denotes the mDM overlap factor for the $\mathrm{TE}_{011}$ mode in an idealized cavity, while $C_{\mathrm{axion}}^n$ denotes the published axion overlap factor of the relevant TM mode. For an idealized cavity, $C^{(010)}_{\mathrm{axion}}=0.69$ and $C^{(011)}_\mDM=0.22$.

In principle, the bound in Eq.~\eqref{eq:coup} should be set at a mass $m_\phi$ determined by the actual TE$_{011}$ resonant frequency of the cavity, which would vary as the cavity is tuned.  In practice, existing cavity experiments do not publish this frequency as a function of their tuning (nor is their tuning mechanism optimized to vary the TE$_{011}$ frequency).  In order to project sensitivities, we instead define a mass mapping
\begin{equation}
m_{\phi}=\frac{f_{011}^{\mathrm{TE}}}{2f_n^{\mathrm{TM}}}\,m_a,
\label{eq:freqmap}
\end{equation}
where $m_a$ varies over the published range of axion masses which a given experiment probes, and $f_n^{\mathrm{TM}}$ and $f_{011}^{\mathrm{TE}}$ are TM and TE frequencies computed for an idealized geometry with the published dimensions of the cavity.  This defines an approximate mass range over which the experiment may be able to scan, although achieving such a range may require different tuning mechanisms than what are currently used.

\begin{table*}[t]
\centering
\scriptsize
\setlength{\tabcolsep}{3.2pt}
\caption{Experimental inputs associated with the AxionLimits data files used in the haloscope recast \cite{MiguelGAlbert}, the data is taken from  \cite{axionlimits} The volume, quality factor, magnetic field, frequency, and $C_a$ columns refer to the published axion-sensitive configuration. The corresponding mDM mass coordinate is obtained from Eq.~\eqref{eq:freqmap}. The listed $C_a$ values are axion-mode form factors and must not be identified with the mDM overlap factors. In the geometry column, CC denotes a circular cylinder and RB a rectangular box; all dimensions are given in metres.}
\label{tab:haloscope-inputs}
\resizebox{\textwidth}{!}{%
\begin{tabular}{llccccccc}
\hline
AxionLimits file & Cavity geometry [m] & $V$ [L] & Reported $Q$ &
$B$ [T] & $\nu$ [GHz] & Axion mode & $C_a$ & mDM mode \\
\hline
\texttt{ADMX} & CC, $a=0.21$ & n.s. & n.s. & 7.6 & 0.466--0.879 & TM$_{010}$ & 0.69 & TE$_{011}$ \\
\texttt{ADMX2018} & CC, $a\simeq0.21$ & 136 & $Q_L\simeq50{,}000$ & 6.8 & 0.645--0.676 & TM$_{010}$ & 0.40 & TE$_{011}$ \\
\texttt{ADMX2019\_1} & CC, $a\simeq0.21$ & 136 & $Q_L\simeq30{,}000$ & 7.6 & 0.677--0.747 & TM$_{010}$ & 0.40 & TE$_{011}$ \\
\texttt{ADMX2019\_2} & CC, $a\simeq0.21$ & 136 & $Q_L\simeq30{,}000$ & 7.6 & 0.760--0.800 & TM$_{010}$ & 0.40 & TE$_{011}$ \\
\texttt{ADMX2021} & CC, $a\simeq0.21$ & 136 & $Q_L\simeq80{,}000$ & 7.5 & 0.801--1.019 & TM$_{010}$ & 0.40 & TE$_{011}$ \\
\texttt{ADMX2024} & CC, $a\simeq0.21$ & 106 & $Q_L\simeq45{,}000$ & 7.6 & 0.792--0.808 & TM$_{010}$ & 0.40 & TE$_{011}$ \\
\texttt{ADMX2025} & CC, $a=0.2095$ & 106 & $Q_L\simeq45{,}000$ & 7.6 & 1.090--1.315 & TM$_{010}$ & 0.40 & TE$_{011}$ \\
\texttt{ADMX\_Sidecar} & CC, $a=0.032$ & 0.38 & $Q_L=2{,}200$--$6{,}200$ & 0.78--3.11 & \begin{tabular}{@{}c@{}}4.20--4.25; 5.09--5.80;\\ 7.17--7.20\end{tabular} & TM$_{010}$ & 0.04--0.61 & TE$_{011}$ \\
\texttt{ADMX\_Sidecar\_AC} & CC, $a=0.032$ & 0.38 & $Q_L=2{,}300$--$6{,}200$ & 3.11 & 4.20--4.25; 7.17--7.20 & TM$_{010}$ & 0.04--0.49 & TE$_{011}$ \\
\texttt{CAPP-1} & CC, $\varnothing0.134\times0.246$
& 3.47 & $Q_L\simeq3.0\times10^4$ & 7.3 & 1.60--1.65
& QTM$_{010}$ & 0.52 & TE$_{011}$ \\
\texttt{CAPP-6} & CC (CAPP-12TB)
& 36.85 & $Q_L=(3.5$--$3.8)\times10^4$ & 10.31 & 1.09--1.11
& TM$_{010}$-like & 0.60 & TE$_{011}$ \\
\texttt{CAPP-9} & CC, $\varnothing0.078\times0.300$
& 1.43 & $Q_L\simeq3.7\times10^4$ & 9.0 & 5.17--5.27
& TM$_{020}$ & $\simeq0.50$ & TE$_{011}$ \\
\texttt{CAPP-MAX} & CC, $\varnothing0.274\times0.640$
& 37.0 & $Q_0=(0.9$--$1.2)\times10^5$ & 10.55 & 1.025--1.185
& TM$_{010}$-like & $\simeq0.60$ & TE$_{011}$ \\
\texttt{HAYSTAC\_PhaseI} & CC, $a=0.051$ & 2.0 & $Q_L\simeq10{,}000$ & 9.0 & 5.60--5.80 & TM$_{010}$ & $\simeq0.50$ & TE$_{011}$ \\
\texttt{HAYSTAC\_PhaseII\_ab} & CC, $a=0.051$ & 1.545 & $Q_L=5{,}915$--$6{,}180$ & 8.0 & \begin{tabular}{@{}c@{}}4.100--4.178;\\ 4.459--4.523\end{tabular} & TM$_{010}$ & 0.43 & TE$_{011}$ \\
\texttt{HAYSTAC\_PhaseII\_cd} & CC, $a=0.051$ & 1.545 & $Q_L\simeq4{,}600$ & 8.0 & \begin{tabular}{@{}c@{}}4.178--4.459;\\ 4.523--4.707\end{tabular} & TM$_{010}$ & 0.43 & TE$_{011}$ \\
\texttt{QUAX2019} & CC, $a=0.013$ & 0.036 & $Q_L=2.01\times10^5$ & 2.0 & 9.067 & TM$_{010}$ & 0.589 & TE$_{011}$ \\
\texttt{QUAX2021} & CC, $a=0.011$ & 0.081 & $Q_L\simeq36{,}000$ & 8.1 & 10.4018 & TM$_{010}$ & 0.69 & TE$_{011}$ \\
\texttt{QUAX2024} & CC, $a=0.0135$ & 0.139 & $Q_L=32{,}200$--$33{,}700$ & 8.0 & 8.83177--8.83777 & TM$_{010}$ & $\simeq0.69$ & TE$_{011}$ \\
\texttt{ORGAN-1a} & CC, $\varnothing0.032\times0.080$
& 0.048 & $Q_L\simeq3500$ & 11.5 & 15.28--16.23
& TM$_{010}$ & $\simeq0.40$ & TE$_{011}$ \\
\texttt{ORGAN-1b} & RB,
$a=0.0056$--$0.0060$, $b=0.0271$, $d=0.0773$
& 0.0118--0.0123 & $Q_L\simeq3700$ & 11.5 & 25.97--27.07
& TM$_{110}$ & $\simeq0.40$ & TE$_{201}$ \\
\hline
\end{tabular}%
}
\end{table*}

To connect this estimate with an experimental benchmark, consider a sub-eV mass near $2\times10^{-6}\,\mathrm{eV}$, corresponding to half the resonant frequency of the TE$_{011}$ cavity mode for the ADMX-like benchmark geometry. Taking $g_{a\gamma}=0.36$, $a=0.21\,\mathrm{m}$ and length $d=1.014\,\mathrm{m}$, one obtains a characteristic fractional millicharge of order $\epsilon_m \sim 10^{-19}$. Using the coupling relation in Eq.~\eqref{eq:coup} together with the frequency map in Eq.~\eqref{eq:freqmap}, previously reported axion-haloscope results from ADMX, CAPP, HAYSTAC, QUAX, and ORGAN, including the ADMX main-cavity and Sidecar runs, the CAPP data sets listed in Table~\ref{tab:haloscope-inputs}, the HAYSTAC Phase I and Phase II scans, the QUAX-$a\gamma$ cavity searches, and ORGAN Phases~1a and 1b \cite{asztalos2010,du2018,boutan2018,braine2020,bartram2021,bartram2023,goodman2025,carosi2025,lee2020capp1,yi2023capp6,bae2024capp9,ahn2024cappmax,brubaker2017,zhong2018,backes2021,jewell2023,bai2025,alesini2019quax,alesini2021quax,alesini2022quax,divora2023quax,rettaroli2024quaxlnf,quiskamp2022organ,quiskamp2024organ}, can be recast in the mDM scenario. For QUAX, only the QUAX-$a\gamma$ resonant-cavity limits on the axion--photon coupling $g_{a\gamma\gamma}$ are used in Fig.~\ref{fig:mdm-power}. Representative experimental inputs for these data sets are summarized in Table~\ref{tab:haloscope-inputs}. Figure~\ref{fig:mdm-power} shows this reinterpretation in the $(m_\phi,\epsilon_m)$ plane. The vertical bands therefore represent a haloscope sensitivity inferred from existing axion data rather than a dedicated direct search for mDM. Some cosmological bounds are also included in Fig.~\ref{fig:mdm-power}. These bounds, as well as the haloscope sensitivity, may depend on the millicharged dark-matter density $\rho_\mDM$ relative to the total dark-matter density $\rho_\DM$.

\section{Astrophysical and Cosmological Constraints}
\label{sec:constraints}
The constraints shown in Fig.~\ref{fig:mdm-power} arise from several different physical effects.  It is useful to explain their origin before discussing how they change when the millicharged component accounts for only a fraction of the total DM.

The pulsar and fast-radio-burst (FRB) constraint of Ref.~\cite{caputo2019} uses the
fact that a population of millicharged particles behaves as a very dilute
electromagnetic plasma.  Radio waves propagate more slowly through a plasma
at lower frequencies, so pulses emitted simultaneously at the source arrive
at different times.  This is the same basic effect used to measure the
ordinary free-electron dispersion measure, but the mDM component gives an
additional contribution,
\begin{equation}
\propto m_e\left(\frac{\epsilon_m}{m_\phi}\right)^2
 \int_{\rm l.o.s.}\rho_{\mDM}(l)\,dl,
 \label{eq:mdm-dispersion-measure}
\end{equation}
where the integral follows the line of sight from the source to the observer.
In physical terms, a larger charge makes the mDM plasma respond more strongly
to the electromagnetic wave, whereas a larger particle mass makes it harder
to accelerate and reduces the delay.  The mDM contribution has the same
leading frequency dependence as ordinary electron dispersion and therefore
cannot generally be separated using one radio-frequency band alone.  The
bound is obtained by comparing the dispersion inferred from radio timing with
an independent estimate of the electron column density, for example from
optical observations.  It is applicable when the radio photons probe the
nonrelativistic plasma response assumed in Ref.~\cite{caputo2019}.

The red-giant constraint of Ref.~\cite{davidson2000} is a stellar-cooling
bound.  Photons propagating through the dense stellar plasma acquire an
effective mass and may be treated as quasiparticles called plasmons.  If
kinematically allowed, a plasmon can decay into a pair of millicharged
particles.  Particles with sufficiently weak interactions escape the star and
carry away energy, opening a cooling channel that is absent from standard
stellar-evolution calculations.  Excessive cooling would delay helium
ignition and alter the observed evolution and number counts of red giants in
globular clusters.  Agreement between these observations and standard stellar
models therefore limits the millicharge.  In the light-mass, free-streaming
regime, Ref.~\cite{davidson2000} finds a nearly mass-independent limit of order
\(\epsilon_m\lesssim 2\times10^{-14}\).  The precise numerical coefficient is
particle-model dependent; in particular, the result quoted in that reference
is calculated for millicharged fermions and its application to a scalar should
be interpreted accordingly.

We also include a constraint on the mDM charge from cosmological bounds on the DM-proton cross-section.  Near the time of recombination, the DM and baryon fluids possess a relative bulk velocity.  Scattering between DM and protons allows transfer of heat and momentum, which can affect observables in the cosmic microwave background (CMB).  Ref.~\cite{Boddy_2018} placed a constraint on Coulomb-like DM-proton scattering based on Planck 2015 data, and Ref.~\cite{Bogorad_2021} translated this into a constraint on the mDM charge.

Finally, we note the existence of astrophysical constraints on mDM arising from interactions with magnetic fields in the interstellar medium~\cite{stebbins2019} and galaxy clusters~\cite{kadota2016}.  In the former scenario, mDM can exchange angular momentum with the Galactic disk, causing the Galaxy to spin down.  In the latter, the trajectories of mDM in galaxy clusters can be dominantly influenced by magnetic fields, rather than gravity, leading to DM density profiles which contradict observations.  Both of these arguments set strong constraints on the DM charge-to-mass-ratio, $\epsilon\lesssim10^{-13\pm1}\cdot(m/\mathrm{GeV})$.  However, we note that both rely on a particle-like treatment of mDM, which may not apply at the low masses considered in this work, where mDM behaves as a wave.  Therefore, we do not show these constraints in Fig.~\ref{fig:mdm-power}.

Let us now discuss how all these constraints change when mDM is a subfraction of DM. Let
\begin{equation}
 f_{\mDM}\equiv\frac{\rho_{\mDM}}{\rho_{\DM}}
\end{equation}
denote the fraction of DM in the millicharged component.  The relevant fraction is not always the same physical quantity: a cavity probes the local
Galactic density and pulsar timing probes a line-of-sight column density.  The cosmological value of \(f_{\mDM}\) relevant for DM-proton scattering need not equal these local fractions in a model where electromagnetic forces modify the spatial distribution of mDM.

The pulsar/FRB and resonant-cavity limits can be grouped
together over the range in which their standard interpretations remain valid.
For radio dispersion, Eq.~\eqref{eq:mdm-dispersion-measure} scales as
\(\propto\epsilon_m^2\rho_{\mDM}\).  For the cavity,
Eq.~\eqref{eq:PmDM} gives
\(P_{\mDM}\propto\epsilon_m^4\rho_{\mDM}^2\).  Holding the corresponding observable at its limiting
value therefore gives the common charge scaling
\begin{equation}
 \epsilon_m^{\rm lim}(f_{\mDM})
 \simeq \epsilon_m^{\rm lim}(1)f_{\mDM}^{-1/2}.
 \label{eq:common-fmdm-scaling}
\end{equation}
Thus, decreasing the mDM abundance weakens both limits by moving their
charge boundary upward.

The red-giant limit belongs to a second category.  The millicharged particles
responsible for stellar cooling are produced thermally inside the star, so
their emission rate does not depend on the pre-existing Galactic or
cosmological mDM density.  Consequently, the red-giant bound is approximately
unchanged when \(f_{\mDM}\) is reduced, provided the same particle physics and
free-streaming assumptions remain valid.

The constraint on the DM-proton cross-section derived in Ref.~\cite{Boddy_2018} does not exhibit a straightforward scaling with $f_\mathrm{mDM}$.  Nevertheless, as $f_\mathrm{mDM}$ decreases, less DM is able to exchange heat with the baryon fluid, and so the constraint indeed weakens. This constraint disappears entirely for $f_\mathrm{mDM}\lesssim0.2\%$.  We also note that the large-scale magnetic field constraints from Ref.~\cite{stebbins2019,kadota2016} are alleviated for sufficiently small mDM subfractions as well.

The local magnetic-deflection condition is also separate from the abundance
rescalings above.  At leading order, it is determined by the fractional
distortion of a given incident wave, so the overall field amplitude cancels
from the ratio of the distorted and undistorted solutions.  Its boundary in
the \((m_\phi,\epsilon_m)\) plane is therefore not shifted by a simple power of
\(f_{\mDM}\).  Nevertheless, the detectable cavity power still decreases with
the local density as described above.  These distinctions are important in
the small-fraction regime: haloscope limits, radio-dispersion limits, stellar
cooling, cosmological limits, and local wave
propagation do not all respond to a reduced mDM abundance in the same way.

Overall, the abundance dependence of the different probes is not universal. Pulsar/FRB dispersion and resonant-cavity limits weaken approximately as $f_\mDM^{-1/2}$, red-giant cooling bounds are essentially unchanged, and constraints from DM-proton scattering eventually disappear. Thus, in the small-fraction regime, haloscope searches can become comparatively more relevant, provided the local abundance of the charged component remains sufficient. The relation between the local fraction $f_\mDM^{\mathrm{loc}}$ that controls the laboratory signal and the cosmological fraction $f_\mDM$ is model-dependent, and should therefore be specified explicitly in any astrophysically consistent charged-DM interpretation.

\section{Conclusion}
\label{sec:conclusion}

We have derived the effective electromagnetic current sourced by coherent scalar mDM in an external vector potential and evaluated its response in a resonant cavity experiment. The resulting source has a different modal structure from the standard axion haloscope signal: in the idealized geometry considered here, the leading overlap is with transverse-electric modes, in particular the TE$_{011}$ mode, rather than with the usual transverse-magnetic TM$_{010}$ mode. This distinction fixes the relevant form factor and the mode-frequency mapping, both of which are essential for a consistent recast of existing axion-haloscope data. In this work, we also derive the conditions under which the deflection of the mDM wave by the cavity's and Earth's magnetic fields can be neglected.

Under ADMX-like benchmark assumptions, the resulting sensitivity can reach fractional charges around $\epsilon_m\sim10^{-19}$ in narrow resonant mass windows, although the interpretation remains conditional on the local mDM abundance, the correct mode overlap and the correct frequency relation. Our results show that resonant cavities can be the strongest laboratory probes of mDM in the $m_\phi\sim\mathrm{\mu eV}$ mass range.  Moreover, in scenarios where mDM constitutes a subcomponent $f_\mDM\lesssim0.2\%$ of DM, they can be the strongest probes overall.  These findings provide motivation for existing cavity experiments to tune and readout the TE$_{011}$ modes of their apparatuses.

\section*{Acknowledgements}
We thank Joerg Jaeckel for discussions on this topic in the early stages of this project.  We also thank Zachary Bogorad and Gordan Krnjaic for discussions on relevant mDM constraints.

This work was initiated as the master's thesis project of M.G. in the MasterCosmosBCN Postgraduate Program in High Energy Physics, Astrophysics \& Cosmology through Universitat Aut\`onoma de Barcelona (UAB) and Institut de F\'isica d’Altes Energies (IFAE).

IFAE is partially funded by the CERCA program of the Generalitat de Catalunya.  S.K. is supported by ERC grant ERC-2024-SYG 101167211 and is funded by the European Union.  Views and opinions expressed are however those of the author only and do not necessarily reflect those of the European Union or the European Research Council Executive Agency.  Neither the European Union nor the granting authority can be held responsible for them.

Fermilab is administered by Fermi Forward Discovery Group, LLC under Contract No. 89243024CSC000002 with the U.S. Department of Energy, Office of Science, Office of High Energy Physics.

\appendix
\clearpage
\onecolumn
\counterwithout*{equation}{section}

\setcounter{section}{0}
\setcounter{equation}{0}
\renewcommand{\theequation}{A.\arabic{equation}}
\renewcommand{\theHequation}{appendix.\arabic{equation}}
\setlength{\mathindent}{2em}
\newcommand{\A}{\mathbf A}
\newcommand{\hphi}{\hat\varphi}
\renewcommand{\hm}{\mathbf{\hat m}}
\newcommand{\hk}{\mathbf{\hat k}}
\newcommand{\hq}{\mathbf{\hat q}}
\newcommand{\hr}{\mathbf{\hat r}}
\renewcommand{\k}{\mathbf k}
\newcommand{\m}{\mathbf m}
\newcommand{\q}{\mathbf q}
\newcommand{\x}{\mathbf x}

\begin{center}
{\Large Appendix: Magnetic Distortion of mDM Background\par}
\end{center}

In the main text, we assumed that the mDM takes the simple form given in Eq.~\eqref{eq:phi_cosine}.  This is the case for a charge neutral background far from electromagnetic (EM) fields.  However, the effect described in this work requires the presence of local EM fields, and so we ought to understand if these fields can distort the mDM background.  In particular, there are two sources of potential concern: the local background magnetic field of the cavity experiment; and large-scale astrophysical magnetic fields, such as the geomagnetic field.  In both cases, we can model the background magnetic field as a dipole (far from the experiment) and compute its effect on mDM via Eq.~\eqref{eq:eom}.  A similar calculation is performed in Ref.~\cite{snipe_hunt_mdm}, focusing on the limit $kr\ll1$, where $k\sim m_\phi v_\DM$ is the DM momentum and $r$ the distance from the center of the dipole.  In our case, we will see that the experimental magnetic field satisfies this condition, although the Earth's magnetic field does not.  We will compute the effect in both limits and understand when the effect is large enough for the mDM distribution to deviate from Eq.~\eqref{eq:phi_cosine}.

Consider a mDM distribution whose evolution is governed by Eq.~\eqref{eq:eom} and takes the form in Eq.~\eqref{eq:phi_cosine} far from the center of the dipole magnetic field.  We write the vector potential of this field as
\begin{equation}\label{eq:dipole}
\A(\x)=\frac{\m\times\x}{4\pi|\x|^3},
\end{equation}
with magnetic moment $\m=-4\pi B_0r_0^3\hm$ (so that the magnetic field at $r=R_0$ along the equator of the dipole is $B_0$).  The form in Eq.~\eqref{eq:phi_cosine} consists of two counter propagating plane waves.  Since Eq.~\eqref{eq:eom} is linear, for simplicity, we may consider just a single plane wave.  That is, we assume the time dependence $\phi\sim e^{-i\omega t}$, with $\omega^2=k^2+m_\phi^2$, so that Eq.~\eqref{eq:eom} becomes%
\footnote{Throughout this appendix, we decompose all vectors into magnitude and direction as $\mathbf v=v\mathbf{\hat v}$.  The one exception is $\x=r\hr$.}
\begin{align}\label{eq:eom2}
(\nabla^2+k^2)\phi(\x)&=V(\x)\phi(\x),\\
V(\x)&=2ie_m\A(\x)\cdot\nabla+e_m^2|\A(\x)|^2.
\end{align}
We are interested in the scenario where the effect of the dipole field is small, so we may employ the Born approximation, namely that
\begin{equation}
\phi(\x)=\phi_0e^{i\k\cdot\x}+\phi_1(\x)
\end{equation}
with $\phi_1(\x)\ll\phi_0e^{i\k\cdot\x}$.  In this case, Eq.~\eqref{eq:eom2} becomes
\begin{equation}
(\nabla^2+k^2)\phi_1(\x)=V(\x)\phi_0e^{i\k\cdot\x},
\end{equation}
which is solved by
\begin{equation}\label{eq:phi1}
\phi_1(\x)=-\frac{\phi_0}{4\pi}\int d^3\x'\,\frac{e^{ik|\x-\x'|}}{|\x-\x'|}\left(-2e_m\A(\x')\cdot\k+e_m^2|\A(\x')|^2\right)e^{i\k\cdot\x'}.
\end{equation}
The integral in Eq.~\eqref{eq:phi1} has two relevant regimes: $kr\equiv k|\x|\ll1$ and $kr\gg1$.  For $m_\phi\sim3\,\mu\mathrm{eV}$ and $v_\DM\sim10^{-3}$, we have $k^{-1}\sim70\,\mathrm m$.  The length scales relevant for the experimental magnetic field ($r\sim\mathrm m$) lie in the $kr\ll1$, while the length scales relevant for the geomagnetic field ($r\sim10^3\,\mathrm{km}$) lie in the $kr\gg1$ regime. Therefore, we will need to evaluate this integral in both limits.

Let us begin with the first case, where the region of interest is much smaller than the mDM de Broglie wavelength.  In this case, the exponents in Eq.~\eqref{eq:phi1} can be ignored.  We will evaluate each term of Eq.~\eqref{eq:phi1} separately, starting with the first term.  For the purposes of evaluating the integral, we must establish a spherical coordinate system.  Let us orient the vertical axis parallel to $\x$ and the $\varphi=0$ plane to contain $\hk\times\hm$.  We denote the length of $\hk\times\hm$ by $\sin\Theta$ and its polar angle by $\alpha$.  In this coordinate system, we describe the vector $\x'$ by spherical coordinates $(r',\theta_1,\varphi_1)$.  In summary, these angles are defined by
\begin{align}
\hk\cdot\hm&=\cos\Theta\\
\hr\cdot(\hk\times\hm)&=\sin\Theta\cos\alpha\\
\hr\cdot\hr'&=\cos\theta_1\\
\hr'\cdot(\hr\times(\hk\times\hm))&=\sin\Theta\sin\alpha\sin\varphi_1,
\end{align}
where $\hr\equiv\frac\x{|\x|}\equiv\frac\x r$ and $\hr'\equiv\frac{\x'}{|\x'|}\equiv\frac{\x'}{r'}$.  In this coordinate system, the first term in Eq.~\eqref{eq:phi1} can be written as
\begin{align}
\phi_1(\x)&\supset-\frac{e_mkB_0r_0^3\phi_0}{2\pi}\int r'^2dr'd(\cos\theta_1)d\varphi_1\cdot\frac1{|\x-\x'|}\cdot\frac{\hr'\cdot(\hk\times\hm)}{r'^2}\\
&=-\frac{e_mkB_0r_0^3\phi_0}{2\pi}\int r'^2dr'd(\cos\theta_1)d\varphi_1\cdot\frac1{\sqrt{r^2+r'^2-2rr'\cos\theta_1}}\cdot\frac{\sin\Theta\left(\cos\alpha\cos\theta_1+\sin\alpha\sin\theta_1\cos\varphi_1\right)}{r'^2}\\
&=-e_mkB_0r_0^3\phi_0\sin\Theta\cos\alpha\int dr'd(\cos\theta_1)\cdot\frac{\cos\theta_1}{\sqrt{r^2+r'^2-2rr'\cos\theta_1}}\\
&=-e_mkB_0r_0^3\phi_0\sin\Theta\cos\alpha\int dr'\cdot\frac{2\cdot\min(r^3,r'^3)}{3r^2r'^2}=-e_mkB_0r_0^3\phi_0[\hr\cdot(\hk\times\hm)].
\end{align}

Now let us address the second term in Eq.~\eqref{eq:phi1}.  To evaluate this term in the integral, it will be beneficial to use a slightly different coordinate system.  Namely, we will now orient the $\varphi=0$ plane to contain $\hm$.  In this coordinate system, we will denote the polar angle of $\hm$ by $\beta$ and describe $\x'$ by spherical coordinates $(r',\theta_1,\varphi_2)$.  We now have the definitions
\begin{align}
\hr\cdot\hm&=\cos\beta\\
\hr\cdot\hr'&=\cos\theta_1\\
\hr'\cdot(\hr\times\hm)&=\sin\beta\sin\varphi_2.
\end{align}
In this coordinate system, the second term in Eq.~\eqref{eq:phi1} becomes
\begin{align}
\phi_1(\x)&\supset-\frac{e_m^2B_0^2r_0^6\phi_0}{4\pi}\int r'^2dr'd(\cos\theta_1)d\varphi_2\cdot\frac1{|\x-\x'|}\cdot\frac{|\hm\times\hr'|^2}{r'^4}\\
&=-\frac{e_m^2B_0^2r_0^6\phi_0}{4\pi}\int\frac{dr'd(\cos\theta_1)d\varphi_2}{\sqrt{r^2+r'^2-2rr'\cos\theta_1}}\cdot\frac{(\sin\theta_1\sin\varphi_2)^2+(\cos\theta_1\sin\beta-\sin\theta_1\cos\beta\cos\varphi_2)^2}{r'^2}\\
&=-\frac{e_m^2B_0^2r_0^6\phi_0}4\int\frac{dr'd(\cos\theta_1)}{\sqrt{r^2+r'^2-2rr'\cos\theta_1}}\cdot\frac{1+\cos^2\beta+(1-3\cos^2\beta)\cos^2\theta_1}{r'^2}\\
&=-\frac{e_m^2B_0^2r_0^6\phi_0}4\int\frac{dr'}{r'^2}\left\{\begin{array}{cl}\frac8{3r}+\frac{4r'^2}{15r^3}-\frac{4r'^2}{5r^3}\cos^2\beta,&r'<r\\\frac8{3r'}+\frac{4r^2}{15r'^3}-\frac{4r^2}{5r'^3}\cos^2\beta,&r'>r\end{array}\right.
\label{eq:divergent}\end{align}
This expression nominally diverges.  This is due to the fact the magnetic dipole potential in Eq.~\eqref{eq:dipole} diverges as $\A(\x')\sim1/r'^2$.  In reality, any dipole field is regulated inside a certain radius.  Let us suppose that Eq.~\eqref{eq:dipole} only applies for $r\gtrsim r_0$.  If we are interested in the distortion of the mDM field at $r\sim r_0$, then the $r'<r$ portion of the integral in Eq.~\eqref{eq:divergent} will be accordingly regulated.  We can see from the general form of Eq.~\eqref{eq:divergent} that the integral will evaluate to something of the form
\begin{equation}
\phi_1(r\sim r_0)\supset-\frac{e_m^2B_0^2r_0^4\phi_0}4\left(c_1+c_2[\hr\cdot\hm]^2\right),
\end{equation}
where $c_1,c_2$ depend on the internal structure of the dipole field.

In summary, we find that, in the limit $kr_0\ll1$, the distortion of the mDM is roughly given by
\begin{equation}
\frac{\phi_1(r\sim r_0)}{\phi_0}\sim-e_mkB_0r_0^3[\hr\cdot(\hk\times\hm)]-\frac{e_m^2B_0^2r_0^4}4\left(c_1+c_2[\hr\cdot\hm]^2\right).
\end{equation}
Defining $\varepsilon\equiv e_mB_0r_0^2$, we see that the two contributions to the distortion are proportional to $kr_0\varepsilon$ and $\varepsilon^2$, respectively.  In other words, in order for the distortion to be negligible, we require $\varepsilon\ll1$.%
\footnote{Ref.~\cite{arza2025} performed a similar calculation, using a delta function vector potential.  They found the distortion to be controlled by the parameters $\kappa\sim\varepsilon/kr_0$ and $\varepsilon$.  This would imply a much more stringent constraint $\kappa\ll1$.  However, a delta function vector potential is unphysical.  Actual magnetic fields are better described by the dipole vector potential given in Eq.~\eqref{eq:dipole}.}
For an experimental magnetic field with $B_0\sim7\,\mathrm T$ and $r_0\sim0.2\,\mathrm m$, this is satisfied for $e_m\lesssim7\times10^{-16}$.

Now let us consider the case $kr_0\gg1$.  In this case, the second term in Eq.~\eqref{eq:phi1} will be negligible.  Moreover, the integral will be dominated by $r'\ll r$.  We can therefore approximate
\begin{equation}\label{eq:largek}
\phi_1(\x)=-\frac{e_mkB_0r_0^3\phi_0}{2\pi}\int d^3\x'\cdot\frac{e^{ik(r-\hr\cdot\x')}}r\cdot\frac{\hr'\cdot(\hk\times\hm)}{r'^2}\cdot e^{i\k\cdot\x'}.
\end{equation}
In order to evaluate this integral, let us choose spherical coordinates with $\q=\k-k\hr$ aligned along the vertical axis and $\hk\times\hm$ in the $\varphi=0$ plane.  Let the polar angle of $\hk\times\hm$ be $\gamma$ and let $\x'$ be described by spherical coordinates $(r',\theta_2,\varphi_3)$, so that
\begin{align}
\hq\cdot(\hk\times\hm)&=\sin\Theta\cos\gamma\\
\hq\cdot\hr'&=\cos\theta_2\\
\hr'\cdot(\hq\times(\hk\times\hm))&=\sin\Theta\sin\gamma\sin\varphi_3
\end{align}
In these coordinates, Eq.~\eqref{eq:largek} evaluates to
\begin{align}
\phi_1(\x)&=-e_mkB_0r_0^3\phi_0\frac{e^{ikr}}{2\pi r}\int r'^2dr'd(\cos\theta_2)d\varphi_3\cdot e^{iqr'\cos\theta_2}\cdot\frac{\sin\Theta\left(\cos\gamma\cos\theta_2+\sin\gamma\sin\theta_2\cos\varphi_3\right)}{r'^2}\\
&=-e_mkB_0r_0^3\phi_0\sin\Theta\cos\gamma\frac{e^{ikr}}r\int dr'd(\cos\theta_2)\cdot e^{iqr'\cos\theta_2}\cos\theta_2\\
&=-2ie_mkB_0r_0^3\phi_0\sin\Theta\cos\gamma\frac{e^{ikr}}r\int dr'\,\frac{\sin qr'-qr'\cos qr'}{q^2r'^2}
\label{eq:divergent2}\end{align}

The integral in Eq.~\eqref{eq:divergent2} is dominated by $r'\ll1/q$.  Since $1/q\ll r_0$, as mentioned before, Eq.~\eqref{eq:dipole} should be regulated at such small scales.  In principle, the exact form of the magnetic field for $r\ll r_0$ can strongly affect the resulting distortion.  We would like to derive model-independent bounds on the distortion, so let us consider the case of a general $\A(\x)$, which takes the form in Eq.~\eqref{eq:dipole} at large $r$.  We can write Eq.~\eqref{eq:largek} in the more general form
\begin{equation}\label{eq:generalA}
\phi_1(\x)=e_mk\phi_0\frac{e^{ikr}}{2\pi r}\int d^3\x'\cdot[\hk\cdot\A(\x')]\cdot e^{i\q\cdot\x'}.
\end{equation}
Let us denote the direction of $\hq$ by $x$, and the other two directions by $y$ and $z$.  Then we may break up the full integral in Eq.~\eqref{eq:generalA} into wavelength-sized integrals along the $x$-direction, that is,
\begin{align}
\phi_1(\x)&=e_mk\phi_0\frac{e^{ikr}}{2\pi r}\int dydz\sum_n\int_{x_n-\pi/q}^{x_n+\pi/q}dx\cdot[\hk\cdot\A(\x')]\cdot e^{iqx}\\
&\approx e_mk\phi_0\frac{e^{ikr}}{2\pi r}\int dydz\sum_n\int_{x_n-\pi/q}^{x_n+\pi/q}dx\left(a_n^{(0)}(y,z)+(x-x_n)a_n^{(1)}(y,z)\right)e^{iqx},
\label{eq:taylor}\end{align}
where $x_n=2\pi n/q$, and in the second line, we have Taylor expanded $\hk\cdot\A$ along the $x$-direction, in terms of
\begin{equation}
a_n^{(m)}(y,z)=\hk\cdot\partial^m_x\A(x_n,y,z).
\end{equation}
Evaluating the $x$-integrals in Eq.~\eqref{eq:taylor} gives
\begin{equation}\label{eq:riemann}
\phi_1(\x)\approx e_mk\phi_0\frac{e^{ikr}}{2\pi r}\int dydz\sum_n\frac{2i\pi}{q^2}a_n^{(1)}(y,z).
\end{equation}
Now the sum in Eq.~\eqref{eq:riemann} can be understood as a Riemann sum.  When interpretted this way, $\phi_1(\x)$ can be transformed back into an integral of the form
\begin{align}\label{eq:riemann_approx}
\phi_1(\x)&\approx ie_mk\phi_0\frac{e^{ikr}}{2\pi qr}\int dydz\int dx\,[\hk\cdot\partial_x\A(x,y,z)]\\&=ie_mk\phi_0\frac{e^{ikr}}{2\pi qr}\int dydz\,[\hk\cdot\A(\infty,y,z)-\hk\cdot\A(-\infty,y,z)]=0
\end{align}
because $\A(\x)$ decays to zero far from the origin.

If the Riemann integral approximation were exactly valid, the distortion would vanish.  However, this approximation is not exact.  For a function $g(u)$ evaluated over an interval $u\in[a,b]$, the difference between the central value and average value over the integral can be roughly bounded by Taylor expanding $g(u)$ over the interval
\begin{equation}
\left|g\left(\frac{a+b}2\right)-\frac1{b-a}\int_a^bdu\,g(u)\right|\lesssim\left|\frac1{b-a}\int_a^bdu\,\frac{\left(u-\frac{a+b}2\right)^2}2g''\left(\frac{a+b}2\right)\right|=\frac{(b-a)^2}{24}\left|g''\left(\frac{a+b}2\right)\right|.
\end{equation}
Applying this logic to the approximation in Eq.~\eqref{eq:riemann_approx}, we find that the distortion should be bounded by
\begin{equation}
|\phi_1(\x)|\lesssim\frac{e_mk\phi_0}{2\pi r}\int dydz\sum_n\frac{\pi^3}{3q^4}\left|a_n^{(3)}(y,z)\right|\approx\frac{\pi e_mk\phi_0}{12q^3r}\int d^3\x'\left|\partial_x^3\A(\x')\right|.
\end{equation}
Assuming that the vector potential only varies on lengthscales comparable to $r_0$,%
\footnote{For instance, we expect $\A(\x)$ to begin deviating from a dipole inside the Earth's core, and the radius of the Earth's core is approximately half of $r_0$.}
this gives a bound on the distortion
\begin{equation}
\left|\frac{\phi_1(r\sim r_0)}{\phi_0}\right|\lesssim\frac{e_mB_0}{k^2}\sim\frac\varepsilon{(kr_0)^2}.
\end{equation}
For $B_0\sim30\,\mu\mathrm T$ and $v_\mathrm{DM}\sim10^{-3}$, requiring the distortion from the Earth's field to be small gives the conservative constraint
\begin{equation}
e_m\lesssim10^{-15}\cdot\left(\frac{m_\phi}{3\,\mu\mathrm{eV}}\right)^2.
\end{equation}

If we assume a more particular form of the vector potential, we can actually achieve a much stronger bound on the distortion.  Let us assume that $\A(\x)$ exhibits the same angular dependence as in Eq.~\eqref{eq:dipole}, but with a more general radial dependence, that is,
\begin{equation}
\A(\x)=\frac{\m\times\hr}{4\pi}f(r).
\end{equation}
We know that at large radii $\A(\x)$ should recover the form in Eq.~\eqref{eq:dipole}, so $f(r)\approx1/r^2$ for large $r$.  Meanwhile, we know that the Earth's magnetic field is constant in the inner core, so $f(r)\propto r$ for small $r$.  In this case, Eq.~\eqref{eq:divergent2} will take the form
\begin{equation}\label{eq:divergent3}
\phi_1(\x)=-2ie_mkB_0r_0^3\phi_0\sin\Theta\cos\gamma\frac{e^{ikr}}{q^2r}\int dr'(\sin qr'-qr'\cos qr')f(r').
\end{equation}
Following the same logic as above, we can approximate the integral in Eq.~\eqref{eq:divergent3} as a sum%
\footnote{In this case, we include the second order term in the Taylor expansion because its integral is $\mathcal O(q^{-2})$ [as is the integral of the first order term].  Further terms will give higher order corrections in $q^{-1}$.}
\begin{align}
\sum_n\int_{r_n-\pi/q}^{r_n+\pi/q}&dr(\sin qr-qr\cos qr)\left[f(r_n)+(r-r_n)f'(r_n)+\frac{(r-r_n)^2}2f''(r_n)\right]\\
&=-\frac{2\pi}{q^2}\sum_n\left[3f'(r_n)+r_nf''(r_n)\right],
\end{align}
where $r_n=(2\pi n+\pi)/q$.  Again, approximating this as a Reimann integral yields
\begin{equation}
-\frac1q\int dr\left(3f'(r)+rf''(r)\right)=-\frac1q\Big[2f(r)+rf'(r)\Big]_{r=0}^\infty=0.
\end{equation}
The error on the Riemann integral approximation is bounded by
\begin{equation}
\frac{\pi^3}{3q^4}\sum_n\left|5f'''(r_n)+r_nf''''(r_n)\right|\approx\frac{\pi^2}{6q^3}\int dr\left|5f'''(r)+rf''''(r)\right|.
\end{equation}
Plugging into Eq.~\eqref{eq:divergent3}, we find the bound
\begin{equation}
\left|\frac{\phi_1(r\sim r_0)}{\phi_0}\right|\lesssim\frac{e_mB_0}{k^4r_0^2}\sim\frac\varepsilon{(kr_0)^4}\implies e_m\lesssim10^{-5}\cdot\left(\frac{m_\phi}{3\,\mu\mathrm{eV}}\right)^4
\end{equation}
for $r_0\sim6400\,\mathrm{km}$.

\clearpage
\twocolumn
\bibliographystyle{JHEP}
\bibliography{references.bib}

\end{document}